\documentclass[12pt]{article}
\usepackage{cite}
\usepackage{graphicx}
\usepackage[normal]{subfigure}
\usepackage{lineno}

\usepackage {color}

\title{A NEW TECHNIQUE FOR DIRECT INVESTIGATION OF DARK MATTER}

\author{R. Bertoni, F. Chignoli, D.Chiesa, M. Clemenza, G. Lucchini,R. Mazza, P. Negri(+),\\ A. Pullia, N. Redaelli, L. Zanotti\\ University  and INFN of Milano Bicocca\\D. Cundy\\ University and INFN of Milano-Bicocca, IFSI Torino and CERN}

\date{7-11-2013}
\begin{document}
\maketitle

%%%%\title{A NEW TECHNIQUE FOR DIRECT INVESTIGATION OF DARK MATTER}
%\author{R.~Bertoni\from{ins:x},F.~Chignoli\from{ins:x},D.~Chiesa\from{ins:x},
%M.~Clemenza\from{ins:x},D.~Cundy\from{ins:y},G.~Lucchini\from{ins:x},R.~Mazza\from{ins:x},
%P.Negri\from{ins:x}$^+$,A.~Pullia\from{ins:x}, N.~Redaelli\from{ins:x},L.~Zanotti\from{ins:x}}
%%%%\author{R. Bertoni,F. Chignoli,D. Chiesa,M. Clemenza,G. Lucchini,R. Mazza,
%%%%P. Negri$^{+}$,A. Pullia,\\N. Redaelli,L. Zanotti}
%%\date{4-11-2013}
%%\inst{ins:x}Universit\`a  e Sezione INFN di Milano Bicocca 
%%\inst{ins:y}Universit\`a di Milano Bicocca;IFSI Torino and CERN\\

%%\address{ Dipartimento di Fisica e Sezione INFN di MILANO BICOCCA}
%%%\author{D. Cundy}
%%\address{Universita' di Milano Bicocca, IFSI Torino and CERN}
$^{+}$ WE WANT DEDICATE THIS PAPER TO HIS MEMORY,WE MISS HIM DEEPLY\\
\vskip 3 mm
%%($^+$) WE WANT DEDICATE THIS PAPER TO HIS MEMORY,WE MISS HIM DEEPLY}\\
%%\maketitle
\bibliography{thebibliography}
\bibliographystyle{plain}
\abstract
{The MOSCAB experiment (Materia OSCura A Bolle) uses a new technique for
Dark Matter search. The Geyser technique is applied to the construction
of a prototype detector with a mass of 0.5 kg and the encouraging results
are reported here; an accent is  placed  on a big detector of 40 kg in 
construction at the Milano-Bicocca University and INFN.\\}

\vskip 1 cm
\section {INTRODUCTION} 
WIMPs (Weak Interacting Massive Particles) are one of the more
suited hypothesis for the non-baryonic candidate for dark matter ; they
indeed satisfy the required density compatible with the cosmological
constraints; they form galactic halos with a  Maxwellian velocity distribution 
around  a mean velocity of  about 230 km/s and with a matter density of
about  0.3 $ GeV/cm^{3}$ at the location of the solar system.\\
The general form of the WIMP interaction with ordinary matter is:\\
$$ \sigma _{A}=4 G_{F}^{2}(\frac{M_{W} M_{A}} {M_{W}+M_{A}})^{2} C_{A}  $$
where $G_{F}$ is the Fermi constant,$M_{W}$ and $M_{A}$ are the mass of the
WIMP and of the target nucleus; $C_{A}$ is an enhancement factor 
which depends on the type of the WIMP interaction. In  Supersymmetry the
spin-independent (SI) or scalar interactions,  proceeds via Higgs or
 squark exchange or both. and $C_{A}$ is given by:\\
$$ C_{A}^{SI}= (1/4\pi)[Z f_{p}+ (A-Z)f_{n}]^{2}$$
$f_{n,p}$ are the WIMP coupling constant to nucleons.\\
On other hand the spin-dependent interaction (SD) with axial-vector coupling,
 involve squarks and Z exchanges and the $C_{A}^{SD}$ is :\\
$$C_{A}^{SD}=(8/\pi)[a_{p} S_{p}+ a_{n} S_{n}]^{2}\frac{J+1}
{J}=(8\pi)(\lambda)^{2}$$ where $S_{p,n}$ are the average spins over
 all protons and neutrons; $a_{p,n}$ are the effective WIMP proton(neutron)
 coupling strengths and J is the total nuclear spin.\\
The enhancement factor is the largest  for nuclei of $^{19}F$.
(see Table 1 and \cite{Ellis})

\begin{center}
\begin{table}[htb]
\caption[bla]{Enhancement factor for SD reactions}
\vskip 0.5 cm
\begin{tabular}{|c|c|c|c|}
Isotope & Spin &Unpaired & $\lambda^{2}$\\
\hline
\hline
$^{7}Li$ &3/2 & p & 0.11\\
$^{19}F$ &1/2 & p &0.863\\
$^{23}Na$ &3/2 & p &0.011\\
$^{29}Si$ &1/2 & n & 0.084\\
$^{73}Ge$ &9/2 & n & 0.0026\\
$^{127}I$ &5/2 & p & 0.0026\\
$^{131}Xe$ &3/2 & n & 0.0147\\
\hline
\end{tabular}
\end{table}
\end{center}

The relation between the  kinetic energy  of the recoiling ions (in the case
of F)and the WIMP's mass is reported in Fig.1 and 
\cite{Levin} where it is shown that to investigate low Wimp Masses 
(around 10 Gev) it is necessary to explore low energy recoils (around 10 keV).\\
In this Figure we have reported indeed the number of expected events per
day and per kg of detector divided by the cross section ($\sigma_{W+F}$ in
pbarn in the case of SD interaction \cite{Marc}).\\
The nuclear form factor of fluorine and also a rough integration on the energy
spectrum of WIMPs are taken into account.\\
%\begin{center}
%$ K(keV)\le 10 KeV\frac{M_{w}}{10 GeV}$\\
%$=10 KeV $ if $ M_{w}=10 GeV$ ;\\
%$=100 KeV$ if $ M_{w}=100(GeV)$;\\ 
%\end{center}
%\vskip 2 mm

\begin{figure}
\vspace*{13pt}

%%\mbox{\epsfig{figure=disegnogeyser.ps,width=10.cm}}
\includegraphics[scale=.5]{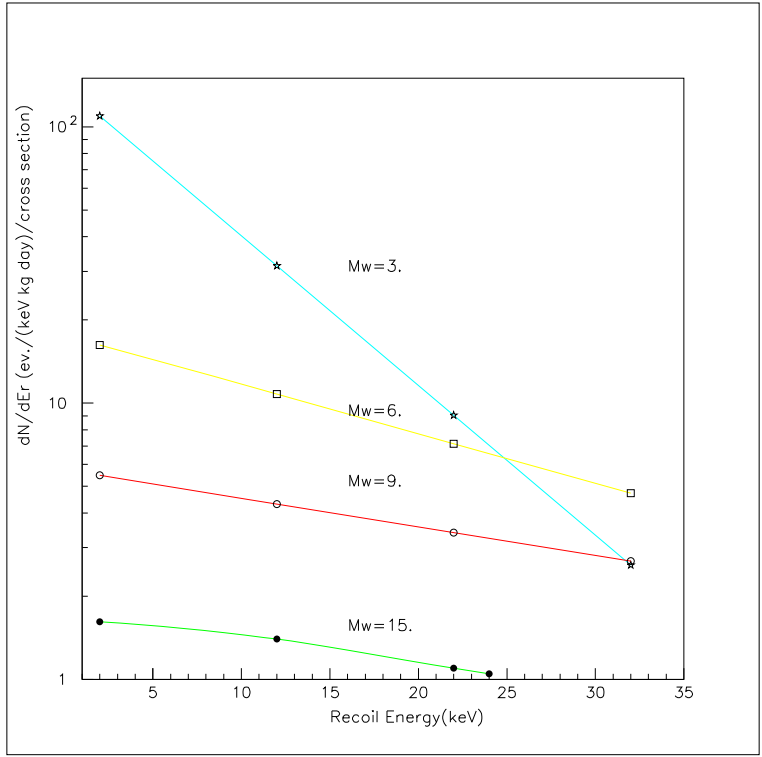}
\vspace*{1.4truein}
\vskip -1.0in
\caption{Kinematics of the elastic scattering of WIMPs  on Fluorine}
\label {Fig. kinematics}
\end{figure}
%%\begin{figure}
%%\vspace*{13pt}

%%\mbox{\epsfig{figure=disegnogeyser.ps,width=10.cm}}
%%\includegraphics[scale=.5]{disegnogeyser.ps}
%%\vspace*{1.4truein}
%%\vskip -1.0in
%%\caption{Sketch of a Vertical section of the Geyser }
%%\label {Fig. 3 Disegno del Geyser}
%%\end{figure}

\begin{figure}
\vspace*{13pt}
%%\mbox{\epsfig{figure=Geyser15.ps,width=10.cm}}
\includegraphics[scale=.5]{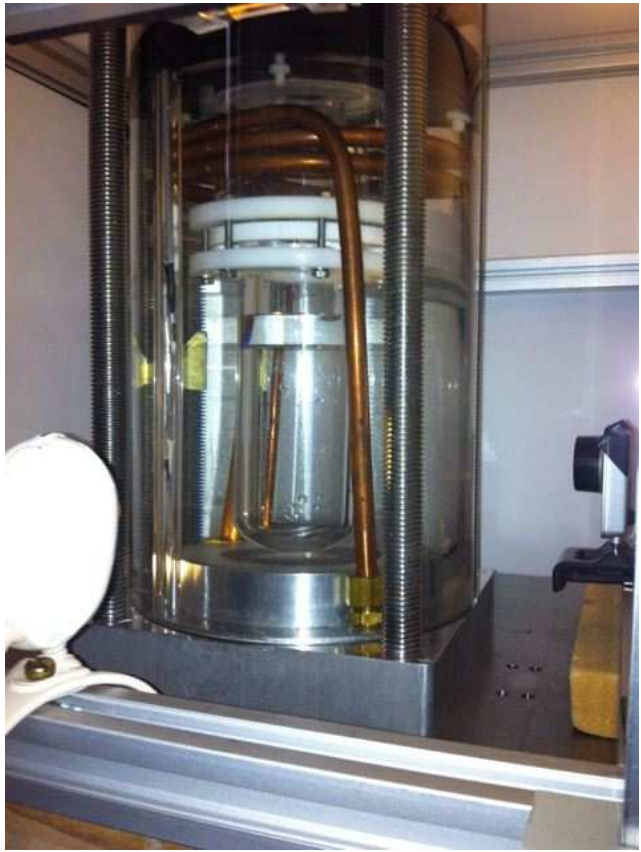}
\vspace*{1.4truein}
\vskip -1.0in
\caption{Internal Geyser's view}
\label {Vista interna del Geyser}
\end{figure}

Many experimental methods have been studied and realized to detect directly
 Dark Matter. In particular we want point out the use of scintillators NaI
\cite{Bernabei}, Liquid Argon \cite{Benetti},
 Xenon \cite{Aprile}, Cryogenic Semiconductors
\cite{Akerib}
and Detectors based on the nucleation of bubbles \cite{B.Hahn},
\cite{COUPP},\cite{Zacek}]. 
In the following we will describe in more detail the technique based on bubble
 formation with the NEW TECHNIQUE of the GEYSER.\\
This kind of detector (Geyser) has never been used for the Elementary 
Particle Physics(it was constructed only once in Bern in 1964 by Hahn and Reist
\cite{Hahn} to detect transuranic nuclei).\\

\section{GENERAL CONSIDERATIONS OF THE NEW TECHNIQUE AND DESCRIPTION OF THE PROTOTYPE}
The technique we have chosen for the direct search of Dark Matter is the
"Geyser Technique" or "Condensation chamber".\
This technique is a variant of the superheated liquid technique of extreme
simplicity.\\
The main volume of the target liquid ($C_{3}F_{8}$ in our case) is kept in
a thermal bath at a constant temperature $T_{L}$.\\
The vapour above the liquid is kept at a temperature $T_{V}<T_{L}$ by
cooling the top of the vessel by a circulating liquid(water).\\
The equilibrium vapour pressure above the liquid is $P_{V}$ so the liquid is in 
a state of under-pressure, and therefore a  superheat of 
 $\Delta p=P_{V}-P_{L}$\,
where $P_{V}=P_{Sat.}(T_{L})$ and $P_{L}=P_{Sat.}(T_{V})$.
A local energy release of energy due for instance to a recoiling ion induced
by a WIMP interaction can produce a vapour bubble which can grow (if over a
threshold in energy) to visible size . This vapour bubble rises in the liquid
and pushs up part of the liquid in the neck(this is the reason of the name 
Geyser).
When equilibrium pressure is reached  the hot
vapour in the top of the vessel recondenses, and the liquid is recovered
into the main volume . The original metastable state returns in a
few seconds  and the system is ready to record a new event. The system
does not require external intervention or recompression .\\

In Fig.2 there is a picture of the apparatus:
in the bottom part of this picture the vessel filled by the liquid freon is 
shown and are also visible the cupper coils that take constant the temperature 
of the liquid  by circulating the water furnished by the thermostates; 
the buffer liquid that separates the freon liquid from the  vapour is Glycol.
In the top part of the apparatus two pressure equalizers are inserted: they are
constituted by two elastic membranes that push the external water
when the pressure of the freon gas become higher and act also in the  reverse
sense. \\
The degree of superheat applied must  exclude the detection of minimum 
ionizing particles (electrons and $\gamma $ rays )  and on the contrary it
must allow the detection with high efficency of the recoiling ions.\\
The principal advantages of the Geyser (and of the  Bubble techniques) are the
 following:\\
 
1)The strong rejection of the particles at minimum ionization (electrons
  and $\gamma $).\\
  
2)The simplicity of the mechanical construction, also for large size
  detectors and therefore low cost.\\
 
3)The very interesting possibility to count multiple neutron interactions and
 hence subtract the neutron background (the interaction length of a neutron is
 of the order of (6-9) cm in our liquid). The double or triple interaction in
 the same frame can be used statistically to evaluate the  number of events
 with a single interaction due to neutrons.\\
 
4)The possibility to distinguish the spin dependent interaction of WIMP
 from spin independent by changing  the liquid used.\\

5)For the Geyser (ONLY) the reset of the detector is automatic and has a very
 short time (few seconds).\\

%%The property 1)  mentioned before is already tested ( Rejection of
%% $\simeq 10^{10}$ of mip particles (see COUPP work) \cite{}\\
A prototype of Geyser has been constructed with a mass of 0.5 kg in Milano-Bicocca \cite{Baudis}.\\
With reference to the Fig.2 the quartz vessel of 0.33 liters is immersed in 
 a water bath and it is surrounded by Cu coils  with an internal
circulating water at the two fixed  temperatures.\\
It contains freon $C_{3}F_{8}$ around 25 C at a pressure of about 6 bar.
The hot freon  is separated from the cold freon vapour  by the neck of the
vessel filled by a buffer liquid (Glycol) with thermal capacity greater than
 that of the water.\\
We would like to point out that in the original Geyser of Hahn no buffer liquid
was used but we found that it improves greatly the stability of the device.\\
The temperature of the two regions of water is kept fixed by two thermostats
 with a precision of 0.1 degrees and the two regions are separated by a loosely fitting
rubber washer.\\
The temperature of the cold vapour was varied from 15-21 C.\\
Everything is surrounded by a cylindrical vessel of plexiglass of thickness 1.5 cm, filled with a
 water/glycol mixture.\\
In order that the flask undergoes only a small over pressure with respect to the water an automatic pressure equalizer using rubber membranes is used.\\
The freon is illuminated by diffuse light, coming from LEDs.\\
To summarize, the Geyser is substancially a Vessel constituted by a ``FLASK'' 
containing the overheated liquid (f.i. some kind of freon) and a "NECK" (containing partially a separation liquid and partially the freon vapour).\\
The scattered ions after an interaction with a neutral particle like a neutron
or a WIMP deposit their energy in very small regions (size of the order
0.05-0.1 micron).\\
In these conditions a bubble can grow and reach a few mm of radius 
(well visible).\\
Two professional digital  cameras monitor in a  continuous way at 50 frames
per second (fps) the volume  in the freon vessel.\\
Some pixels undergo a change of luminosity when a bubble is generated.\\
 At this point a trigger is launched and a stream of pictures  is registered
 (between -50 and + 50 frames starting from the trigger).
%%\begin{figure}
%%\vspace*{13pt}
%%\begin{center}
%%\mbox{\epsfig{figure=allsequence.ps,width=10.cm}}
%%\includegraphics[scale=0.1]{allsequence.ps}
%%\end{center}
%%\vspace*{1.4truein}
%%\vskip -1.0in
%%\caption{ Evolution of a bubble}
%%\label { Growing of the size of a bubble}
%%\end{figure}

After that, the stream of data is stored and visually scanned to see the
 evolution of the bubbles.\\
The bubble reaching the  superior part of the Geyser finds a lower temperature,
becomes again liquid and goes back to the hot region of the overheated
liquid.\\
 This is the fundamental cycle that brings back our Geyser in the initial
 conditions,with a dead time of a few seconds.\\

\section{RESULTS FROM THE PROTOTYPE}

We are working in Milano-Bicocca at the IV floor in  a  Laboratory
provided by the University and INFN.\\
Over the last couple of years we have carried out a large number of runs in
which the temperature of the liquid and vapour  have been varied and  also the 
amount of liquid freon and Glycol.\\
These experiments were carried out in order to find a device that was stable over very long periods
of time,sensitive to Carbon and Fluorine recoils of about
5 keV kinetic energy and insensitive to minimum ionizing particles.\\
Bubble formation is  well understood \cite{Bugg} and depends on the critical radius $R_c =2 \sigma/\Delta p $,
where $\sigma$ is the surface tension of the liquid and $\Delta p$ the pressure difference between the vapour inside the bubble and the liquid.\\
Another important quantity is the critical energy $E_c$ necessary for a visible
 bubble formation.\\
$E_c$ is a function of $ R_c$, $\sigma$, $\Delta p$ and the latent heat of 
evaporation of the liquid.\\
In Fig.3 is shown the energy loss dE/dx for C and F ions and also electrons.\\
Therefore if the energy of recoil is greater than $E_c$( the critical energy)
and stopping power satisfaies the relation $(dE/dx)2 R_c> E_c$, then a bubble 
will form. On Fig.3 we show also several sensitivity zones for various vapour
temperatures and liquid-vapour temperature differences DT; the experimental 
regions in which we must work are reported in squared boxes.\\
 Note that our work region is far away from that for detection of  electrons.\\
In Fig 4 the critical energy is shown as a function of T(vapour) for various
values of DT.\\
We can see that to reach a threshold of 10 keV for the recoiling ions we must 
reach a difference in temperature DT between the liquid and the vapour
$ >7.5 ^{°}C $; to reach a threshold of 3 keV we need a DT of $ 9.0 ^{°}C $.\\

\begin{figure}
\vspace*{13pt}
%%\mbox{\epsfig{figure=disegnogeyser.ps,width=10.cm}}
\includegraphics[scale=.5]{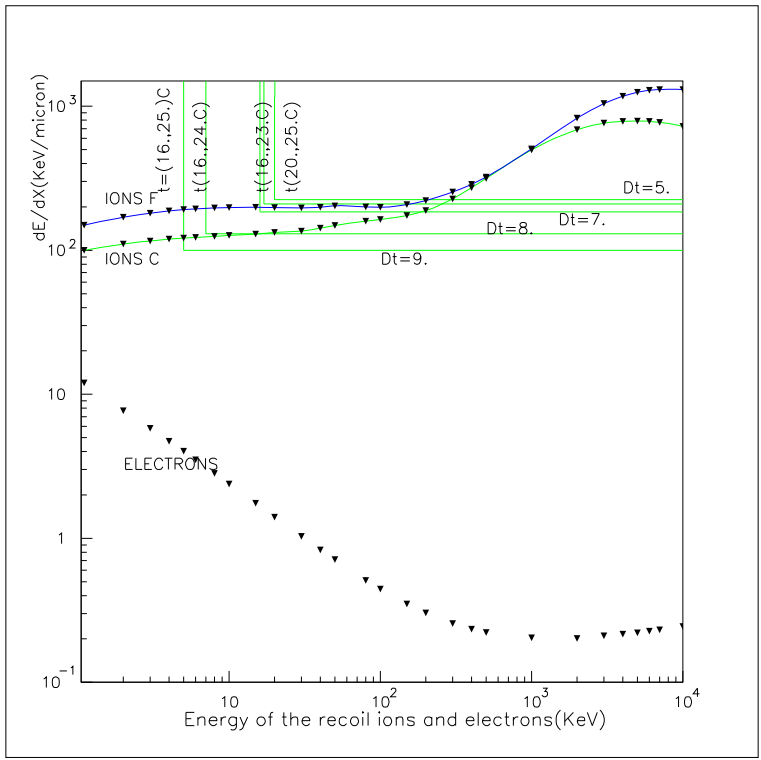}
\vspace*{1.4truein}
\vskip -1.0in
\caption{dE/dx for ions and electrons }
\label {dE/dx}
\end{figure}

\begin{figure}
\vspace*{13pt}
%%\mbox{\epsfig{figure=disegnogeyser.ps,width=10.cm}}
\includegraphics[scale=.5]{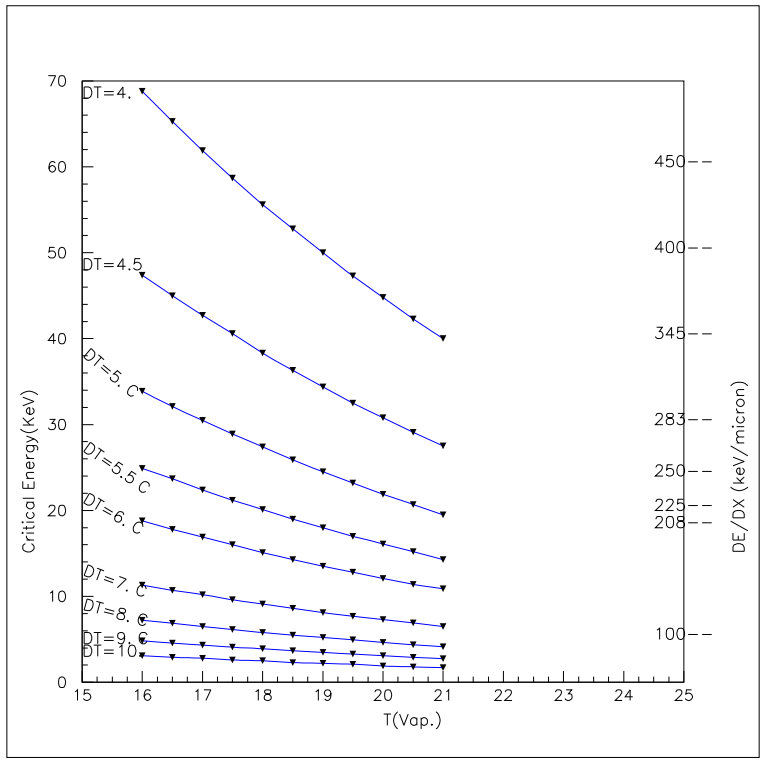}
\vspace*{1.4truein}
\vskip -1.0in
\caption{Critical energy as a function of DT and T(vap) }
\label {Energy threshold as a function of D(T)}
\end{figure}

\begin{figure}
\vspace*{13pt}
%%\mbox{\epsfig{figure=disegnogeyser.ps,width=10.cm}}
\includegraphics[scale=.5]{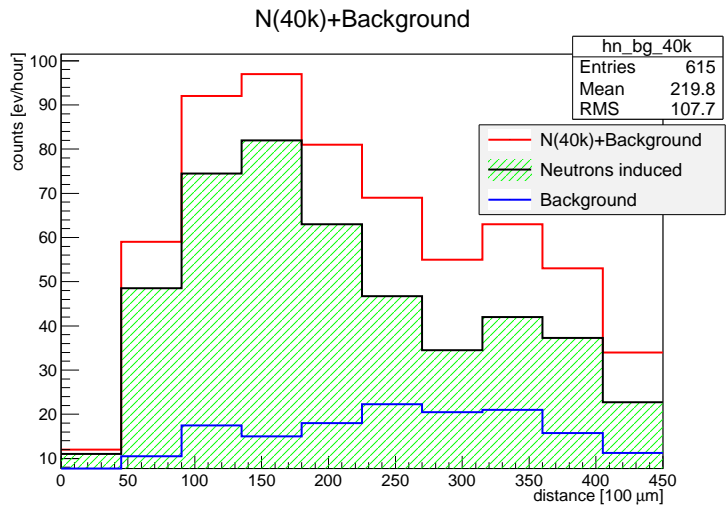}
\vspace*{1.4truein}
\vskip -1.0in
\caption{Background and a Neutron source}
\label {Fig. With a neutron source}
\end{figure}

\begin{figure}
\vspace*{13pt}
%%\mbox{\epsfig{figure=disegnogeyser.ps,width=10.cm}}
\includegraphics[scale=.5]{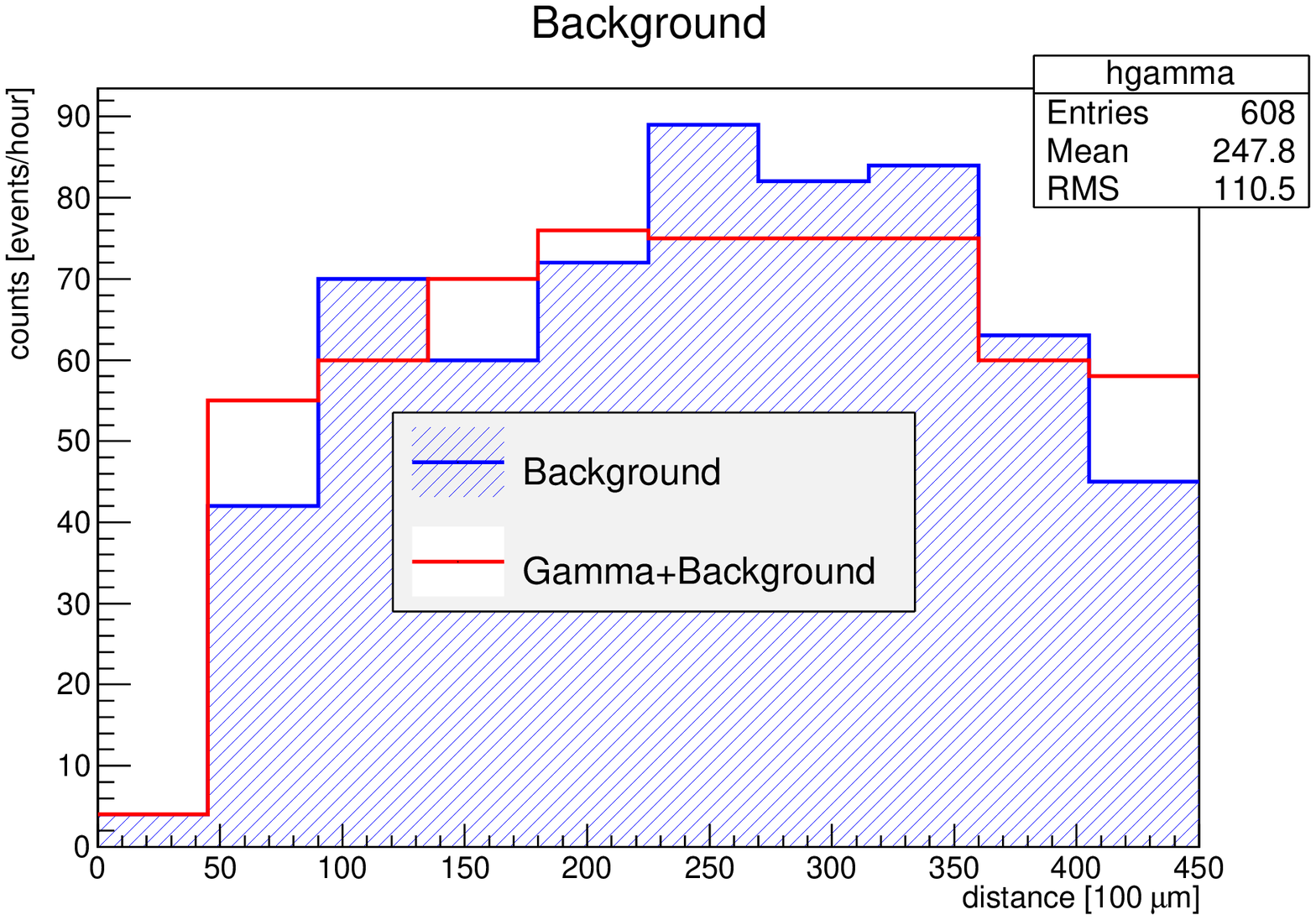}
\vspace*{1.4truein}
\vskip -1.0in
\caption{Backround and a gamma source}
\label {Fig. With a gamma source}
\end{figure}

In the previous chapter  it was said that the characteristic of a Geyser
 must be a high rejection of electrons and $\gamma$ accompanied
by an easy  detection of  nuclear recoils(similar to the recoiling ions
due to an interaction of a WIMP).\\
To test this, we placed outside the detector (at a minimal distance from the
freon) a neutron source ($Am-Be$ -40 kBq).The results are shown in Fig.5 and
we can see that we are very sensitive to the detection of neutrons.\\
 After that we put a gamma rays source (20 kBq $ ^{22}Na$) near the detector
and in Fig.6 are shown the background distribution and that obtained with 
a  Gamma source ($Na^{22}$).\\
We  can remark that in the latter case we obtained compatible results: no 
excess in events in presence of the radiative source! We can hence evaluate
 the rejection factor for electromagnetic showers to be $<10^{-7}$; 
this confirms the COUPP result \cite{Behnke}: rejection factor $<10^{-10}$.\\
By varying the amount of freon in the flask and the height of the GLYCOL we
have managed to obtain extremely stable conditions which allowed a complete
threshold scan above 5 keV, and run for several months.\\
The temperature of the fluid was 25 C and the expected threshold
variation with the vapour temperature is shown in Fig.7.\\

\begin{figure}
\vspace*{13pt}
%%\mbox{\epsfig{figure=curvaseguita.ps,width=10.cm}}
\includegraphics[scale=.5]{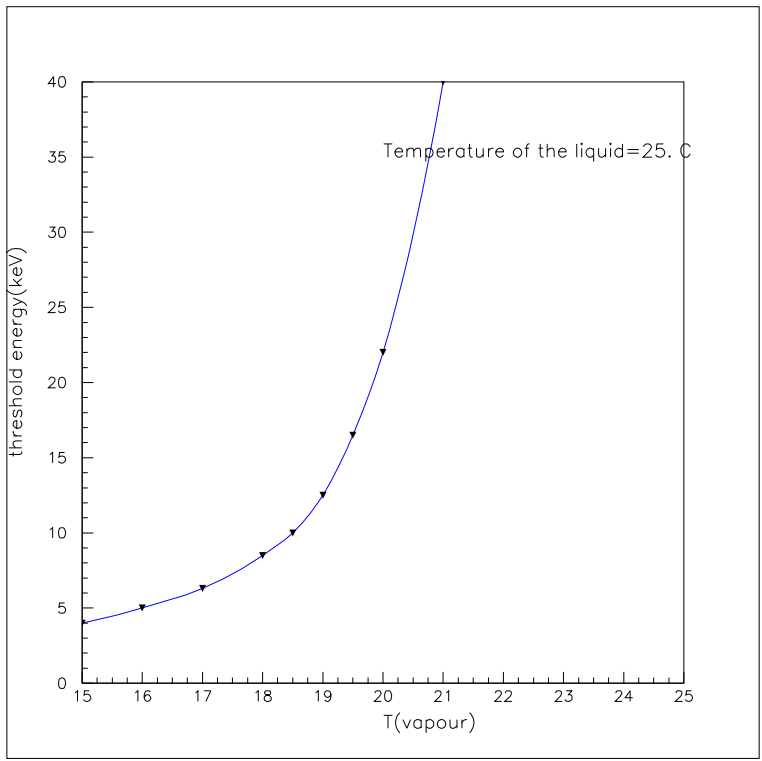}
\vspace*{1.4truein}
\vskip -1.0in
\caption{Followed curve}
\label {Followed curve}
\end{figure}

\begin{figure}
\vspace*{13pt}
%%\mbox{\epsfig{figure=disegnogeyser.ps,width=10.cm}}
\includegraphics[scale=.5]{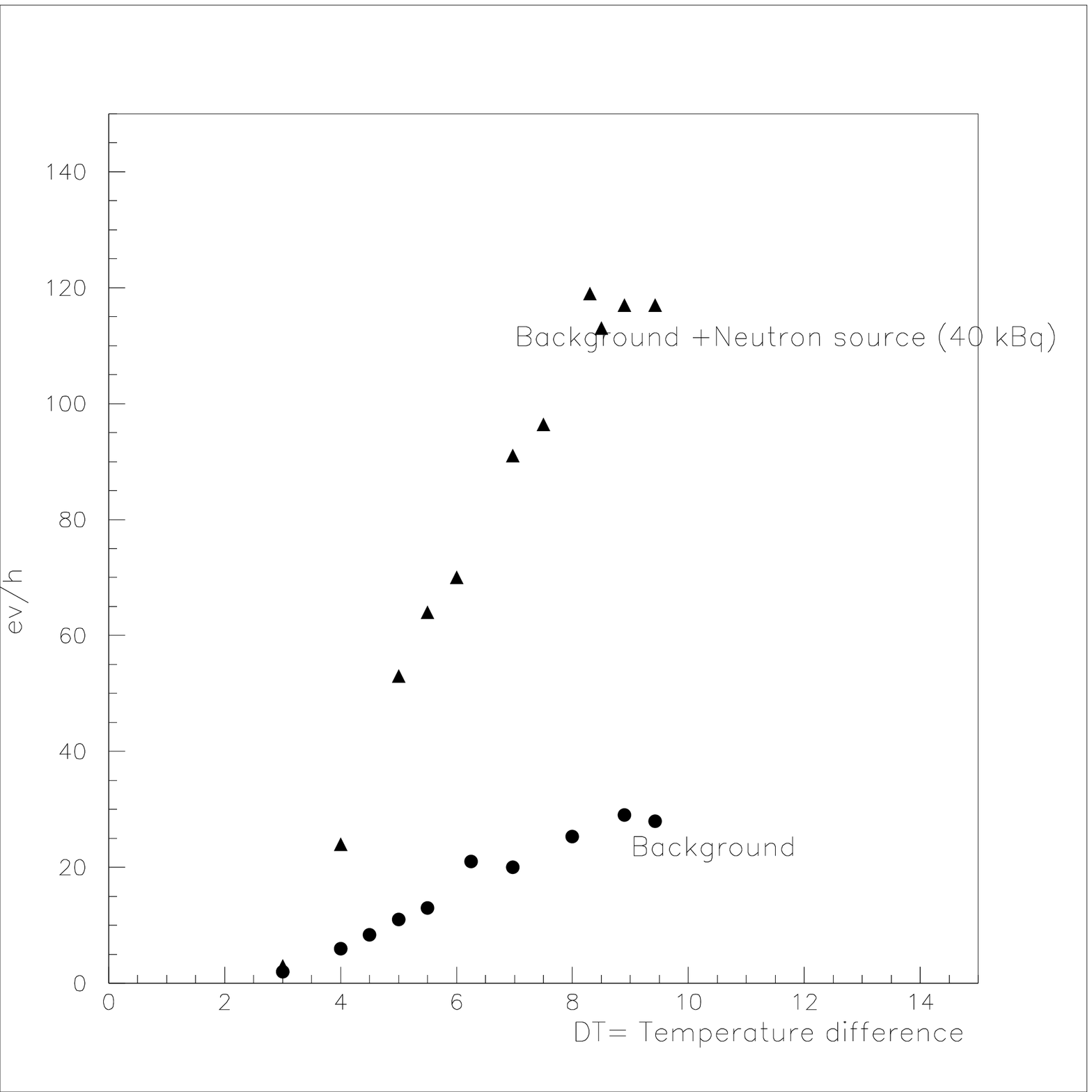}
\vspace*{1.4truein}
\vskip -1.0in
\caption{Recent counts/DT measurements}
\label {Misure Recenti}
\end{figure}

Fig.8 shows the number of events/hour obtained for the background and the
neutron source as a function of DT. An important feature of this cumulative
curve is that a plateau seems to  be reached.\\
In order to compare our data to what is expected from the neutron source we
have performed  Monte-Carlo calculations using the MCNP package coming from
Los Alamos \cite{Manual}. MCNP is a general purpose coupled neutron/photon/electron
Monte Carlo transport code. It is particularly suitable for neutron transport
simulation thanks to the capability to model arbitrary three-dimensional
configuration of material and the continuous-energy cross sections treatment 
used to simulate the transport effects.\\

\begin{figure}
\vspace*{13pt}
%%\mbox{\epsfig{figure=chiesa2.eps,width=10.cm}}
\includegraphics[scale=.5]{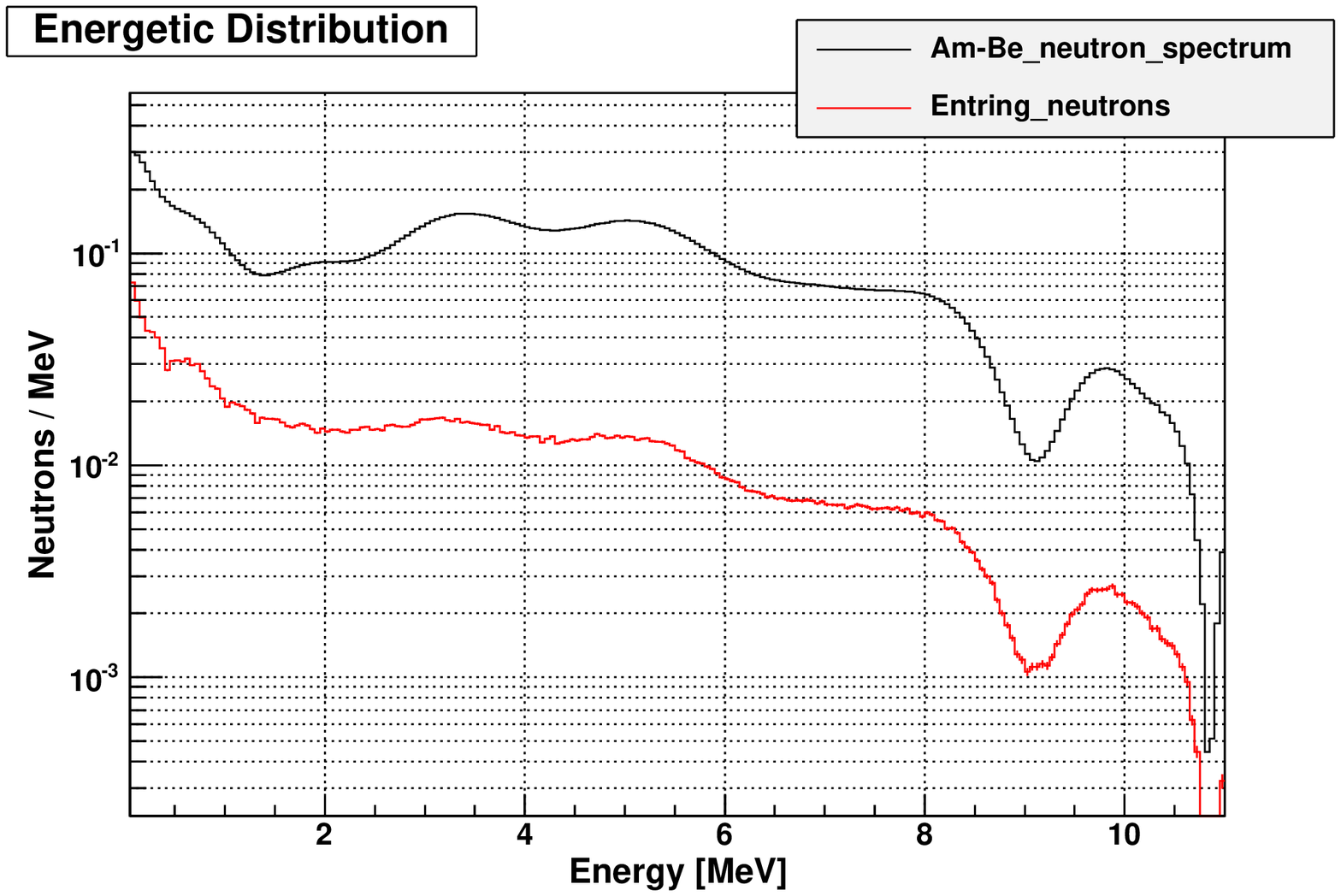}
\vspace*{1.4truein}
\vskip -1.0in
\caption{Neutron spectra from Am-Be source }
\label {Spettri di Am-Be}
\end{figure}

\begin{figure}
\vspace*{13pt}
%%\mbox{\epsfig{figure=chiesa3.eps,width=10.cm}}
\includegraphics[scale=.5]{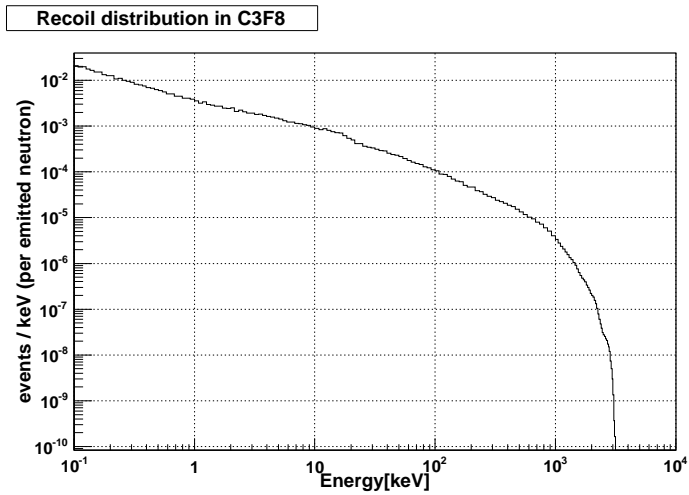}
\vspace*{1.4truein}
\vskip -1.0in
\caption{Monte Carlo calculation }
\label {MC calculation}
\end{figure}

\begin{figure}
\vspace*{13pt}
%%\mbox{\epsfig{figure=disegnogeyser.ps,width=10.cm}}
\includegraphics[scale=.5]{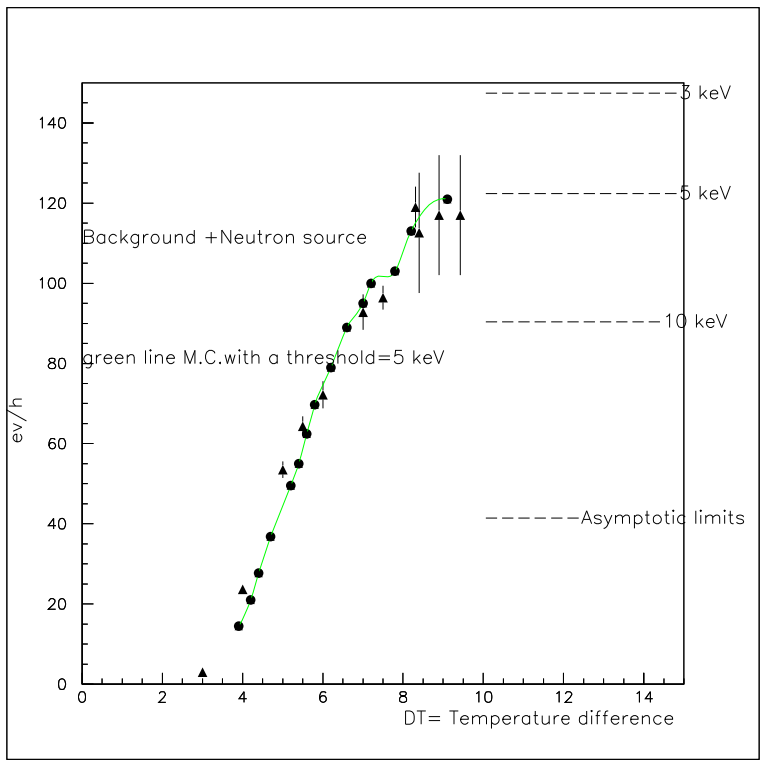}
\vspace*{1.4truein}
\vskip -1.0in
\caption{Comparison with the events-Integrated Distribution and (green line)
(MC + Background}
\label {comparison with the events}
\end{figure}

\begin{figure}
\vspace*{13pt}
%%\mbox{\epsfig{figure=disegnogeyser.ps,width=10.cm}}
\includegraphics[scale=.5]{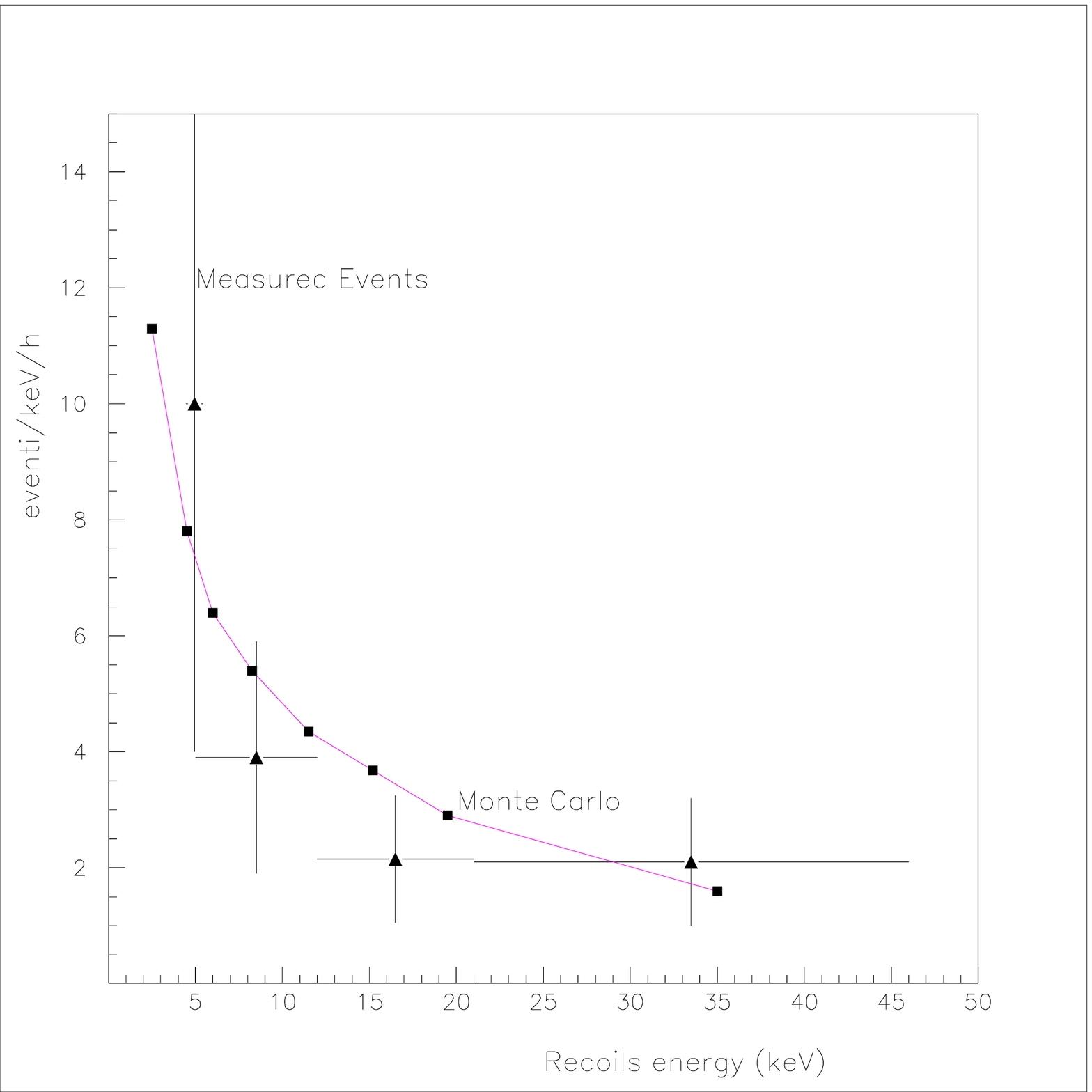}
\vspace*{1.4truein}
\vskip -1.0in
\caption{Comparison with the events-Differential Distribution}
\label {comparison with the events}
\end{figure}

The neutron energy regime is from 5-10 keV to 20 MeV for all isotopes.\\
In Fig.9 the emitted neutron spectrum is shown along with neutron spectrum 
entering the sensitive freon.\\
Fig.10 shows the energy distribution of the recoiling nuclei expected per
emitted neutron.\\
In Fig.11 we compare the distribution (M.C. results + the meaured background)
with the corresponding experimental distribution and we can see a very good
general agreement with a threshold of 5 keV; the reported errors are the 
statistical errors only.\\
We also obtained from our data the differential energy  distribution of the
observed recoils by making:\\
\begin{itemize}
\item a)The background subtraction(difference in rates) at each value of DT.
\item b)The energy distribution of neutrons (background subtracted) 
as a function of the energy threshold and obtained by evaluating the 
differences between contiguous rates of the previous distribution(b).\\ 
\item c)The use of the relation between the Energy treshold and DT shown in
Fig.4.
\end{itemize}

A direct comparison of the M.C. prediction and the measured spectrum of 
neutrons is reported in Fig.12 and a good agreement is obtained also in this
 case.\\
 
\begin{figure}
\vspace*{13pt}
%%\mbox{\epsfig{figure=disegnogeyser.ps,width=10.cm}}
\includegraphics[scale=.5]{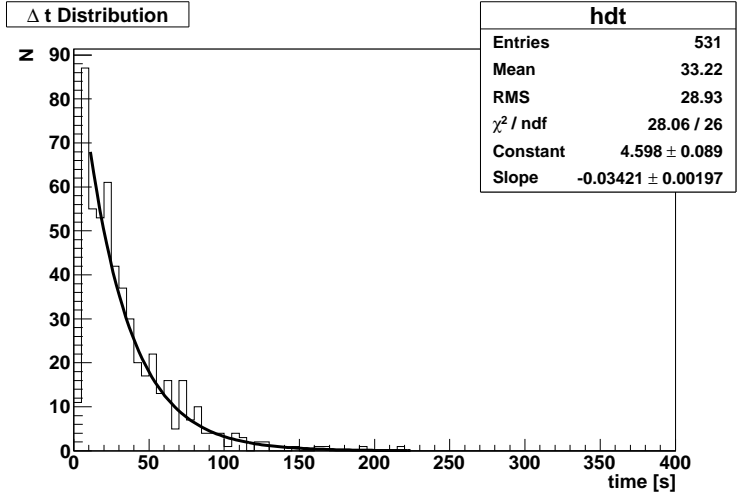}
\vspace*{1.4truein}
\vskip -1.0in
\caption{ Distribution of the time difference between two successive  events and time resolution of the detector(about 5 sec.}
\label {distribuzioni di tempo }
\end{figure}

Another good result we obtained is the following:\\
We have plotted the time difference between successive events and observed 
the expected exponential(Fig.13).\\
It is observed that there is a depletion in the first bin and we conclude that
 the maximum dead time (recovery time) is ~5 seconds.\\

\section{BACKGROUND FOR FUTURE EXPERIMENTS WITH LARGER GEYSERS}
We distinguish two types of problems which would affect the working of larger
Geysers:\\
\underline{ a)NON PARTICLE INDUCED INSTABILITY FOUND IN THE PROTOTYPE}\\
During the long series of tests and  measurements at different temperatures
and different values of DT, we came indeed across two problems:\\

i)Instability induced by the walls of the vessel (some boiling points).
To counteract this effect we decided to cover the internal wall of the vessel 
with a layer of special paint with nanotechnological deposition properties;
after this we have measured with an Atomic Force  Microscope (AFM)
the average dishomogeneity of the wall and the result is $(8.40 \pm 0.40)$ nm.
\\
 This kind of problem was very much reduced and practically disappeared.\\

ii)Instability  from the contact surface between Freon(the sensitive liquid) 
and Glycol ( the buffer liquid).\\
This contact instability has been solved by varying the relative quantity of
liquid freon with respect to glycol.\\
We can remark that these kinds of background (not induced by particles)
in any case can be removed by the definition of a fiducial volume in a big
detector if they are small.

\underline {b)PARTICLE BACKGROUND FOR THE FUTURE DETECTOR OF 40 kg}\\
We are assembling indeed a larger detector of 40 kg ; in that detector we
believe that the main backgrounds in general will be:\\
\begin{itemize}

\item a)The elecron and gamma rays ;\\
 we have seen that a rejection of  our type of detectors  is $\simeq 10^{10}$
 [\cite{Behnke}].\\
 This background is negligible if the freon is produced from a petroleum
 source.\\
 
\item b)The $ \alpha $ decay of impurities in the liquid or in the wall of the
  container vessel;\\
  For this background we are investigating the so called "acoustic trigger" . 
  When a bubble is produced a sound is emitted and the intensity and shape
 of the signal is different in at least two cases:
 an $ \alpha $ decay and a recoil of a nucleus . The range of a recoiling ions
 is  indeed $ <0.1 \mu m $,while the range of an $ \alpha $ particle of 5 MeV
 is of the order of $40 \mu m $ and the length of the signal is longer and 
 stronger.\\ 
  The theory of the sound emission \cite{Martynyuk} in a bubble formation is not well
 developed, but a lot of experience was reached by the experiments for Dark
 Matter search with Superheated Drop Detector (SDD). In 
\cite{Viktor}  a time
 sequence of the bubble's sound emission is reported. In the same 
 Reference the possibility to separate the ion's recoils from the 
 $ \alpha $ decay is shown at the level $10^{-3}$.\\

\item c)Neutrons coming from outside; for this background we plan to count
events with two bubbles, three bubbles etc.(The interaction length of a
neutron is of the  order of (6-9) cm and so they can give multiple
interactions in the liquid Freon).\\
It is possible  after, to estimate the expected number of neutron
interaction with only one bubble. The eventual excess of this kind of events 
could be interpreted as due to WIMPs.\\
\end {itemize}

\begin{figure}
\vspace*{13pt}
%%\mbox{\epsfig{figure=disegnogeyser.ps,width=10.cm}}
\includegraphics[scale=.5]{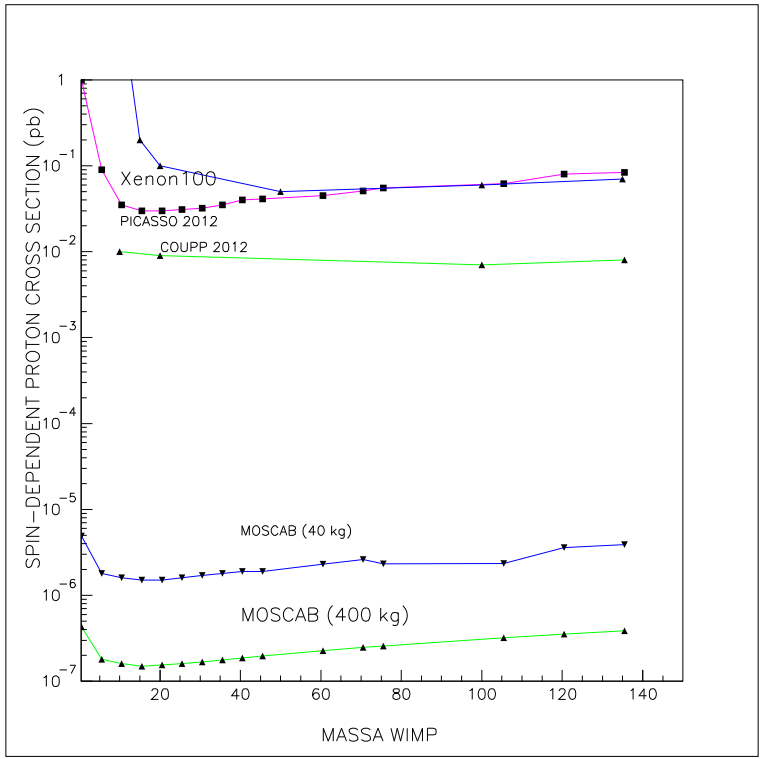}
\vspace*{1.4truein}
\vskip -1.0in
\caption{Expected cross section sensitivity}
\label {Cross section}
\end{figure}

The possible results for our detector are reported in  Fig.14
for two values of the detector mass and in the hypothesis of zero 
background(this aspect of our experiment will be discussed in a future
pubblication):\\
40 Kg (= First module) and 400 Kg (= 10 Modules).\\
We remark that in the SD  case, our sensitivity cold be much better
(by 5 order of magnitude)than that obtained for the results pubblished by PICASSO,COUPP and
Xenon100.\\

\section{Conclusions}
A new technique for the direct investigation of Dark Matter has been developed.
The good results obtained with a Geyser prototype (with a low threshold= ~few
keV) motivate the construction of larger detector of this type and the 
40 kg detector will be ready as soon as possible to obtain very good physics
results at the LNGS.\\
We also would like to claim that this kind of detector would be useful
in a neutrino beam to investigate the interaction $\nu+C=\nu+C$.

\section{Acknowledgements}
We would like to acknowledge the funding authorities of INFN(CSN V) and the
Physics Departement of the Milano-Bicocca University for providing us the
site and many structures related to the experimens.\\
We would like also thanks colleagues that discussed with us many problems
we incountered during the experimens; in particular: A. Ghezzi, P.Dini,
E. Previtali, M. Paganoni, S. Ragazzi, D. Menasce, T. Tabarelli.\\
We thank also Ing. F. Alessandria and Prof. O. Citterio for enlighting 
discussion and help.\\
 Thanks are due also to Prof. A. Borghesi and Dr. Trabattoni  that allowed 
us to use the Atomic Force Microscope and made the relative measurement.

\vskip 3 mm
\end{document}